\documentclass[aps,prx,twocolumn,amsmath,amssymb,longbibliography,superscriptaddress]{revtex4-1}
\usepackage{graphicx}
\DeclareGraphicsExtensions{.pdf,.jpg,.png,.eps}
\usepackage{amsfonts}
\usepackage{amsmath}
\usepackage{amssymb}
\usepackage{color}
\usepackage{enumerate}

\usepackage{dcolumn}
\usepackage{bm}

\begin{document}

\title{Large anomalous Hall, Nernst effect and topological phases in the 3d-4d/5d based \\oxide double perovskites}

\author{Kartik Samanta$^*$}
\affiliation{Max Planck Institute for Chemical Physics of Solids, 01187 Dresden, Germany}
\email{Kartik.Samanta@cpfs.mpg.de}

\author{Jonathan Noky}
\affiliation{Max Planck Institute for Chemical Physics of Solids, 01187 Dresden, Germany}

\author{Iñigo Robredo}
\affiliation{Max Planck Institute for Chemical Physics of Solids, 01187 Dresden, Germany}
\affiliation{Donostia International Physics Center, 20018 Donostia - San Sebastián, Spain}

\author{Juergen Kuebler}
\affiliation{Technische Universitaet Darmstadt, 64289 Darmstadt, Germany}

\author{Maia G. Vergniory$^*$}
\affiliation{Max Planck Institute for Chemical Physics of Solids, 01187 Dresden, Germany}
\affiliation{Donostia International Physics Center, 20018 Donostia - San Sebastián, Spain}
\email{Maia.Vergniory@cpfs.mpg.de}

\author{Claudia Felser$^*$}
\affiliation{Max Planck Institute for Chemical Physics of Solids, 01187 Dresden, Germany}
\email{Claudia.Felser@cpfs.mpg.de}

	\begin{abstract}
		Magnetic topological quantum materials are attracting considerable attention owing to their potential technological applications. However, only a small number of these materials have been experimentally realized, thereby giving rise to the need for new stable magnetic topological quantum materials. Magnetism and spin–orbit coupling, two essential ingredients of the oxide materials, lead to various topological transport phenomena such as the anomalous Hall and anomalous Nernst effects, which can be significantly enhanced by designing an electronic structure with a large Berry curvature.  In that respect, double perovskites with the general formula A$_2$BB$'$O$_6$ with an alternating ordered arrangement of two transition metal sites, B(3d) and B$'$(4d/5d), present attractive possibilities as they are robustly stable against oxidation under ambient conditions and versatile. These double perovskites also offer a high energy scale for magnetism as well as strong spin-orbit coupling with a high magnetic ordering temperature. Here, using first-principles density functional theory calculations, we present a comprehensive study of the intrinsic anomalous transport for 3d-4d/5d based cubic and tetragonal stable double perovskite (DP) compounds. A few of the DPs exhibit a very large anomalous Hall effect with a distinct topological band crossing in the vicinity of the Fermi energy. Our results show the importance of symmetries, particularly the mirror planes, as well as the clean topological band crossing near the Fermi energy, which is primarily contributed by the  5d-t$_{2g}$ for large anomalous Hall and Nernst effects.

\end{abstract}

	\maketitle
	\section{Introduction}
Since the theoretical prediction of topological insulators\cite{Top-insulator-2005,Bernevig-2006,Hassan-Kane-2010,Weider_review} and topological semi-metals \cite{PhysRevX.5.011029,Rev-Yan-2017,RevModPhys.90.015001}, researchers in the field of condensed matter and materials science have extensively endeavored to design, characterize, and synthesize novel topological materials \cite{Rev-Yan-2017,Hassan-Kane-2010,Science-2015,Wieder-2022} considering their potential applicability in the field of quantum information\cite{Kitaev-2003}, coherent spin transport\cite{Nature-Mat-2012}, and high-efficiency catalysis\cite{Adv-Mat-2017}. Recent advances in elementary band representation theory have accelerated the prediction of novel insulating and (semi-)metallic non-magnetic and magnetic topological phases of matter in real materials\cite{Bradly-2017,Cano-2018,Cano-PRB-2017,Maia-2019,Maia-2022,Nature-Mag-topo-2020,Nature-review-Mag-topo-2022}. However, the prediction of new stable magnetic topological materials with large transport characteristics such as like large anomalous Hall conductivity (AHC) and Nernst conductivity (ANC) in strongly correlated magnetic oxides is more challenging and systematic calculations have been conducted only for a few selected strongly correlated oxides \cite{SrRuO3-Science,CrO2-2018,Oxides-2019,Jap-Cry-2020,Jap-strain-2021,Wang-2018,DP-topo1, DP-topo2,Rev-topo-oxide, DP-SFMO}. 
 
In correlated magnetic systems, the interplay between magnetism, crystal symmetry, and spin-orbit coupling (SOC) underlies the various topological transport phenomena and are highly intriguing from an experimental perspective\cite{Burkov-2014,Silvia-2020,Ye-2018}.
In three-dimensional (3D) ferromagnets, spin-polarized bands often display topologically non-trivial nodal points and nodal lines (NLs)\cite{Kim-2018,Hasan-2016,Noky-2019,Noky-2020,Shekhar-2021,nodal2,CoS2-exp}. The lifting of band degeneracy because of the SOC around a nodal structure gives rise to increased Berry curvature around it, leading to a variety of topological transport phenomena, including the anomalous Hall effect (AHE) as well as the anomalous Nernst effect (ANE) \cite{Nagaosa-2010,Noky-2020,Noky-linear-2019,SrRuO3-2d}.

In the case of magnetic topological materials with large anomalous Hall (AHC) and Nernst (ANC) conductivities, it is important to have a magnetic site with a localized magnetic moment, which can provide a high energy scale for magnetism, as well as a magnetic site occupied by a heavy element with strong SOC. From the perspective of designing new magnetic topological materials, it is also important to have materials with physically separated ions that host magnetism and ions that host strong SOC, to avoid the problem resulting from the interplay between the correlation effect and SOC at the same site.
 A promising class of materials in this context is the rocksalt-ordered \cite{Rock-salt} double perovskite (DP) compounds with the general formula A${_2}$BB$'$O${_6}$. These compounds present the possibility of hosting a 3d transition metal (TM) ion at the B site and 4d/5d TM ions at the B$'$ site, offering a suitable playground for exploring exotic properties that depend on the competing energy scales associated with the bandwidth, Coulomb correlation, and SOC for the 3d and 4d/5d orbitals at these two sites. From an experimental point of view, these compounds are easy to produce, are robustly stable against oxidation in ambient air, and possess a high magnetic ordering temperature. However, because most conventional topological materials have low transition temperatures, their application to potential technologies is complicated. Several prominent 3d-4d/5d DP compounds with remarkable properties, such as half-metallicity at room temperature \cite{Half-metal} and colossal magnetoresistance \cite{SFRO-mag} are known. These compounds are excellent candidates for real-life applications, such as spintronics and low dissipation devices\cite{Coey2001}. In addition, in many of these 3d-4d/5d DP materials, the low-energy physics is primarily governed by the t$_{2g}$-orbital bands of the 4d/5d TM ions with strong SOC, with the possibility of a topological band crossing near the Fermi energy with highly efficient transport characteristics, such as AHC and ANC. In spite of the extensive studies of the electronic and magnetic properties of these materials \cite{DP-review,DP-review-TSD,samanta-CaFeCoNiO,samanta-CaCoNiOsO}, the topological characteristics of these materials have not yet been adequately investigated. This motivated our study, in which we considered selected cubic and tetragonal 3d-4d/5d-based experimentally reported double perovskites with the aim to explore and understand new stable topological materials with large AHC and ANC. Our results show the possible existence of a clean topological band crossing near the Fermi energy, primarily contributed by the 5d-t$_{2g}$ orbitals with large anomalous Hall and Nernst conductivities. In certain DP compounds, particularly in Ba$_2$NiReO$_6$ a mirror symmetry protected topological nodal line and their drumhead surface state were observed very close to the Fermi energy. In addition, the AHC and ANC of these compounds are almost three times larger than those of the extensively studied prototype metallic ferromagnetic oxide material SrRuO$_3$ \cite{SrRuO3-Science,SrRuO3-2d}.

\section{Computational details}	

For the calculations, we considered stable cubic and tetragonal 3d-4d/5d-based DP compounds whose structural data was taken from the ICSD database\cite{ICSD} with a 3d magnetic element at the B site and 4d/5d magnetic element at the B$'$ site to enable us to investigate topological transport characteristics such as the anomalous Hall and Nernst effects. We considered the experimentally measured lattice constant, atomic position, and space group as inputs for the density functional theory (DFT) calculations.

The DFT-based plane-wave projected augmented wave (PAW) method, as implemented in the Vienna ab initio simulation package (VASP), was employed \cite{vasp,paw1,paw2}. 
We used a $12\times 12 \times12$ ($12\times 12 \times10$) k-points mesh and a plane wave cut-off of 520\,eV, for the self-consistent calculations of cubic (tegragonal) compounds. These selections of the k-mesh and the plane-wave cut-off were found to result in good convergence of the total energy.
We used the Perdew-Burke-Ernzerhof (PBE) \cite{pbe} exchange correlation functional within the generalized gradient approximation (GGA). 
The electron-electron correlation effects beyond GGA at the strongly correlated 3d and 4d/5d magnetic sites were taken into account using the supplemented on-site Hubbard U correction via the GGA+U method \cite{lda+u}. 
Considering that the B site is occupied by a 3d transition metal and the B$'$ site by a 4d/5d transition metal with wider bands compared to the B site, the correlation effect would be expected to be stronger at the B site compared to the B$'$ site.
Thus, we chose U-values of 4.0 eV and 1.6 eV for the B and B$'$ sites of the considered 3d-4d/5d-based DPs, respectively. The selection of these values was motivated by the U-values used for 3d-4d/5d-based DP compounds in previous studies \cite{Kartik-CaCoNiOsO6,Kartik-FeCoNiOSO6,Samanta-Nadoped,DP-review-TSD}, in which the electronic structure and magnetic ground state were found to be in good agreement with the experimental results. We initially checked our calculations by varying the U-values of a few compounds, but overall, the electronic structure and magnetic ground state remained unaltered. The effect of the SOC is included in the fully relativistic schemes and for all calculations the magnetic moments are chosen to be parallel to the (001) direction.

We calculated the contribution of the intrinsic Berry curvature to the anomalous Hall conductivity by employing the Wannier interpolation technique\cite{wan1}. To compute the Berry curvature, we first constructed a maximally localized Wannier functions (MLWFs) Hamiltonian projected from the GGA+$U$+$SOC$ Bloch wave functions using the Wannier90 tool \cite{wan2,wan3,wan4}. Atomic orbital-like MLWFs of the A$-d$, B$-3d$, B$^{'}-4d/5d$ and O$-p$ states were considered to construct the tight-binding Hamiltonian, which reproduced the spectrum of the system accurately in the energy window of 2.0\,eV around the Fermi energy.
The wannierization was repeated with different inner and outer energy windows until it reproduced the spectrum of the system quite accurately in the energy window of $\pm$ 2.0\,eV around the Fermi energy.

	With the tight-binding model Hamiltonian, we calculated the intrinsic AHC using the linear response Kubo formula approach as follows \cite{Kubo-2004}:
	\begin{eqnarray}
		\Omega^{z}_{n}(\mathbf{k}) = -\hslash^{2} \sum_{n \neq m} \frac{\operatorname{2 Im} \langle u_{n\mathbf{k}}| \hat{ v}_{x}|u_{m\mathbf{k}}\rangle \langle u_{m\mathbf{k}}|\hat{v}_{y}|u_{n\mathbf{k}}\rangle}{(\epsilon_{m\mathbf{k}}-\epsilon_{n\mathbf{k}})^{2}},
	\end{eqnarray}
where $\Omega^{z}_{n}(\mathbf{k})$ is the Berry curvature of band $n$, $\hslash\hat{v}_{i} ={\partial \hat{H}(\mathbf{k})}/{\partial k_{i}} $ is the $i$'th velocity operator, $u_{n\mathbf{k}}$ and $\epsilon_{n\mathbf{k}}$ are the eigenstates and eigenvalues of the Hamiltonian $\hat{H}(\mathbf{k})$, respectively.
	
	Subsequently, we calculated the AHC, given by: 
	\begin{equation}
		\begin{aligned}
			\sigma^{A}_{H}=&  -\frac{e^{2}}{\hbar} \sum_{n} \int_{BZ} \frac{d{\bf k}}{\left(2 \pi\right)^3}f_{n} \Omega^{z}_{n}(\mathbf{k}),
		\end{aligned}
	\label{AHC}
	\end{equation}
and the ANC, $ \alpha^{A}_{H}$, as proposed by Xiao et al. \cite{ANE-prl,ANE-2010}

	\begin{equation}
		\begin{aligned}
			\alpha^{A}_{H}=&  -\frac{1}{T} \frac{e^{2}}{\hbar} \sum_{n} \int_{BZ} \frac{d{\bf k}}{\left(2 \pi\right)^3} \Omega^{z}_{n}(\mathbf{k})\\
			&[(E_{n}-E_{F})f_{n}+k_{B}T ln(1+exp(\frac{E_{n}-E_{F}}{-k_{B}T}))],
		\end{aligned}
	\label{ANE}
	\end{equation}
	
where T is the actual temperature, $f_{n}$ is the Fermi distribution, and E$_{F}$ is the Fermi level. We used a k-point mesh of $300 \times 300 \times 300$ for the calculation of the AHC and ANC using the Eq. \ref{AHC} and Eq. \ref{ANE}, respectively, and  T=300 K for the calculation of the ANC. The surface states were calculated using tight-binding methods by the combination of the Wannier90\cite{wan2,wan3,wan4} and WannierTools software packages\cite{WanTool1,WanTool2}.		

\section{Result and discussion}
	
	\begin{figure*}[ht!]
		\centering
			\includegraphics[width=0.8\textwidth]{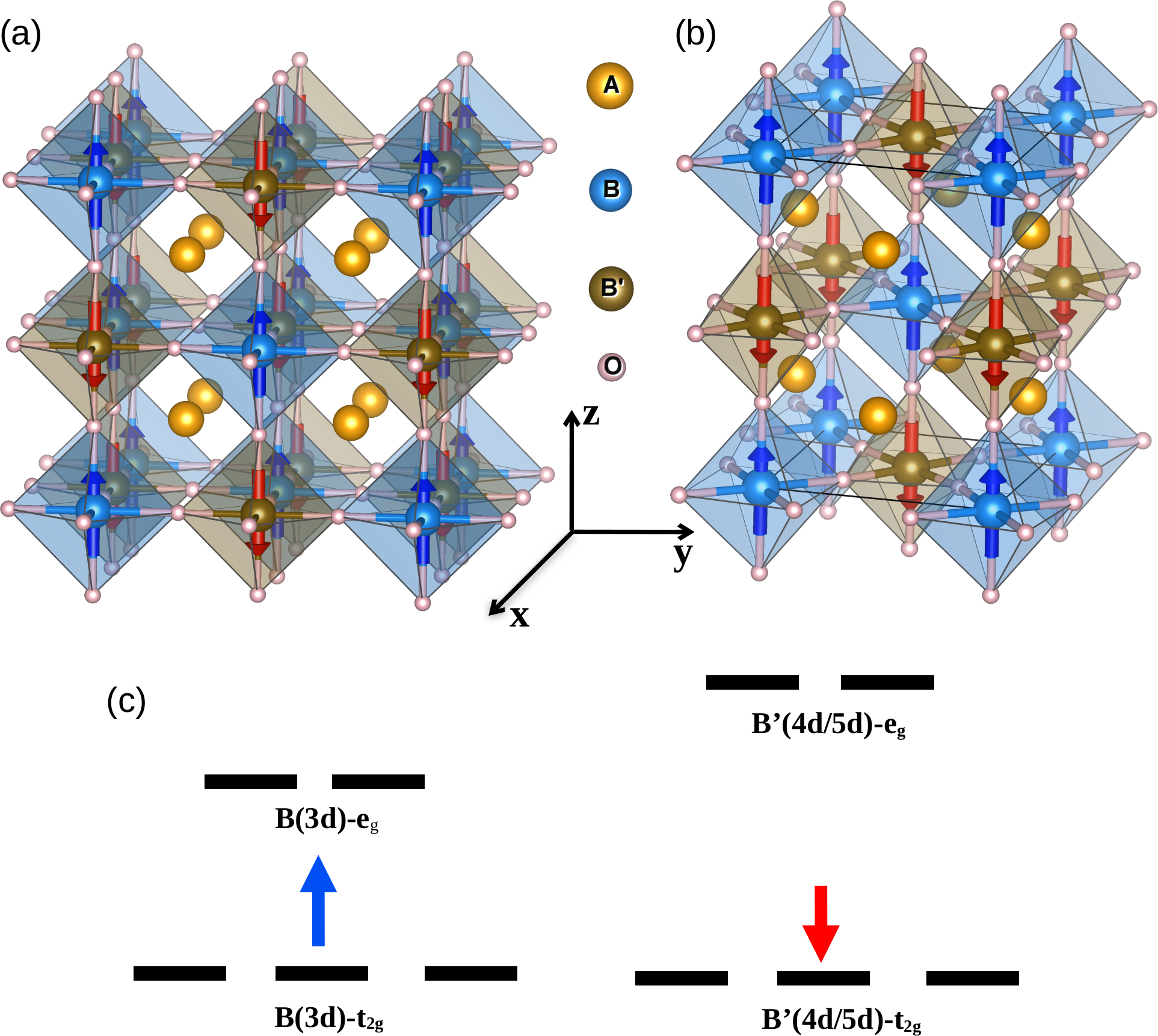}
		\caption{Crystal structure of (a) cubic space group Fm-3m (No. 225) and (b) tetragonal space group I4/m (No. 87) double perovskites. The arrows indicate the direction of magnetic moment at the B and B$'$ magnetic sites. (c) Schematic energy level diagram of the d-levels in the cubic octahedron crystal field at the 3d-B and 4d/5d B$'$ sites.}
		\label{struc}
	\end{figure*}

\begin{table*}[ht!]
\label{table1}
\centering
\caption{Summary of the results of selected materials considered in this study. Listed are the space group (SG), charge state, and electron occupancy at the B-3d and B$'$-4d/5d sites, magnetic moment at the B-3d and B$'$-4d/5d sites, theoretical magnetization per formula unit (f.u.), AHC, maximum AHC, ANC, and maximum ANC. The maximum values are obtained in an energy window of 200 meV around the Fermi level. $\Delta$E indicates the energy distance in eV of the maximum values with regard to the Fermi level.}
\begin{tabular}{cccccccccccccccccccccccccccc} \toprule
{Material} & {SG} & B-3d & B$'$-4d/5d  && {B} && {B$'$} && {M} &&& {AHC}&&{AHC$_{max}$($\Delta$E)}&&{ANC}&&{ANC$_{max}$($\Delta$E)} \\ 
         &  &site & site  && ($\mu_{B}$)&&($\mu_{B}$)&&($\mu_{B}$/f.u.) &&&  (S cm$^{-1}$) && (S cm$^{-1}$)&& (A m$^{-1}$ K$^{-1}$)&&(A m$^{-1}$ K$^{-1}$) \\
\newline \\
\hline
Ba$_2$FeMoO$_6$& 225 &  Fe$^{3+}$ - 3d$^5$ &Mo$^{5+}$ - 4d$^1$ && 3.971 && -0.399 &&3.99 &&& 6.94 && 37(0.2) && 0.043&& -0.18(0.2) \\
Ba$_2$CoMoO$_6$& 225 &  Co$^{2+}$ - 3d$^7$ &Mo$^{6+}$ - 4d$^0$ && 2.697 && 0.029 && 2.99 &&& 0.0 && 0.00 && 0.00&& 0.00 \\
Ba$_2$MnMoO$_6$& 225 &  Mn$^{2+}$ - 3d$^5$ &Mo$^{6+}$ - 4d$^0$ && 4.537 && 0.109 && 4.99 &&& 0.0 && 0.00 && 0.00&& 0.00 \\
Ba$_2$FeReO$_6$& 225 &  Fe$^{3+}$ - 3d$^5$ &Re$^{5+}$ - 5d$^2$&& 3.897 && -0.920 && 3.06 &&& 33.43 && 329(0.2) && 0.04&& -1.2(0.14) \\
Ba$_2$MnReO$_6$& 225 &  Mn$^{2+}$ - 3d$^5$ &Re$^{6+}$ - 5d$^1$  && 4.516 && -0.6611 &&4.032 &&& 278 && 536(0.03) && 1.52&& -2.1(0.10) \\
Ba$_2$NiReO$_6$& 225 &  Ni$^{2+}$ - 3d$^8$ &Re$^{6+}$ - 5d$^1$ && 1.702 && -0.706 &&1.027 &&& 495 && 571 (0.01) && -1.120&& -2.83(0.05) \\
Ba$_2$NiWO$_6$& 225&  Ni$^{2+}$ - 3d$^8$ &W$^{6+}$ - 5d$^0$  && 1.710 && 0.061 &&2.013 &&& 0.0 && 0.0 && 0.00&& 0.0 \\
Sr$_2$NiMoO$_6$& 87 &  Ni$^{2+}$ - 3d$^8$ &Mo$^{6+}$ - 4d$^0$ && 1.673 && 0.073 &&1.996 &&& 0.00 && 0.0 && 0.00&& 0.00 \\
Sr$_2$FeMoO$_6$& 87 &  Fe$^{3+}$ - 3d$^5$ &Mo$^{5+}$ - 4d$^1$ && 4.058 && -0.554 &&3.99 &&& 73.68 && 99(0.17) && 0.03&& 0.94(0.16) \\
Sr$_2$CoOsO$_6$& 87 &  Co$^{2+}$ - 3d$^7$ &Os$^{6+}$ - 5d$^2$ && 2.678 && -1.183 &&1.155 &&& 300 && 517(0.12) && -1.610&& 3.13(0.2) \\
Sr$_2$CoReO$_6$& 87 &  Co$^{2+}$ - 3d$^7$ &Re$^{6+}$ - 5d$^1$ && 2.444 && -0.438 &&2.027 &&& -22 && 200(0.19) && -0.52&& -1.97(0.2)\\
Sr$_2$CrOsO$_6$& 87 &  Cr$^{3+}$ - 3d$^3$ &Os$^{5+}$ - 4d$^3$ && 2.741 && -1.795 &&0.184 &&& 0.00 && 0.00 && 0.00&& 0.00 \\
Sr$_2$NiOsO$_6$& 87&  Ni$^{2+}$ - 3d$^8$ &Os$^{6+}$ - 5d$^2$  && 1.677 && -0.957 &&0.369 &&& -3.10 && 48 (0.08) && 0.24&& -1.12(0.19) \\
Sr$_2$NiReO$_6$& 87 &  Ni$^{2+}$ - 3d$^8$ &Re$^{6+}$ - 5d$^1$  && 1.671 && -0.649 &&1.031 &&& 447 && 447 && -0.3&& 2.54(0.01) \\  
\hline\hline
\end{tabular}
\end{table*}

In the present study, we systematically investigated 3d-4d/5d-based cubic and tetragonal stable DP compounds in an attempt to identify and further our understanding of new stable magnetic topological quantum materials with large AHE. 
The primary structural building blocks of the considered compounds are the corner-sharing B(3d)-O$_6$ and B$'$(4d/5d)-O$_6$ octahedra (cf. Fig. \ref{struc}(a,b)). As a result of the tetragonal distortion, the octahedra in a tetragonal DP are slightly distorted, whereas they are perfect octahedra in a cubic DP. In the cubic crystal field of regular octahedra, the $d$-orbitals split into a higher energy level of two-fold (four-fold including spin) degenerate e$_g$ orbitals and a lower energy level of three fold (six-fold including spin) degenerate t$_{2g}$ orbitals, as schematically shown in the Fig. \ref{struc}c. The electron-electron correlation is expected to decrease from 3d to 4d to 5d transition metal (TM) elements, whereas the bandwidth and spin-orbit coupling strength increases from 3d to 4d to 5d TM elements owing to the enhanced delocalization of electronic wave functions and the higher atomic number (Z), respectively. Consequently, exchange splitting at the 3d-B site is likely to be quite large. However, crystal field splitting at the 4d/5d-B’ site is expected to be very large relative to the exchange splitting owing to the strong delocalization of the electronic wave function. 

The Large crystal field splitting at the 4d/5d-B$'$ site results in the stabilization of the low spin state, and when the 4d/5d-B$'$ orbital is less than or equal to half filled, the empty e$_g$ orbitals would lie far above the Fermi level, with a negligible impact on the electronic structure and transport properties. The main contributing orbitals around the Fermi energy would be the three 4d/5d-t$_{2g}$ orbitals with strong SOC. In general, the energetically filled 3d orbitals of the B sites, lie below the Fermi energy, which has a finite degree of hybridization with the occupied 4d/5d-t$_{2g}$ orbitals through the O-p orbitals.

 From both an experimental and theoretical perspective, it is important to have a topological material with a clean band structure with a topological crossing near the Fermi energy.  Depending on the electron occupation of the 3d orbitals of the B site (thereby effectively controlling the exchange splitting), which lie below the Fermi energy, we can effectively control the mixing of the occupied 3d orbitals with the partially occupied 4d/5d-t$_{2g}$ states. In the present study, we considered 3d-4d/5d-based cubic and tetragonal DPs with 0, 1, 2, or 3 electrons in the 4d/5d-t$_{2g}$ state near the Fermi energy, which can provide strong SOC and many electrons in the 3d B sites. 
 Table I presents the calculated results of the compounds with a large AHC and/or ANC.

\begin{figure*}[ht!]
		\centering
			\includegraphics[width=0.9\textwidth]{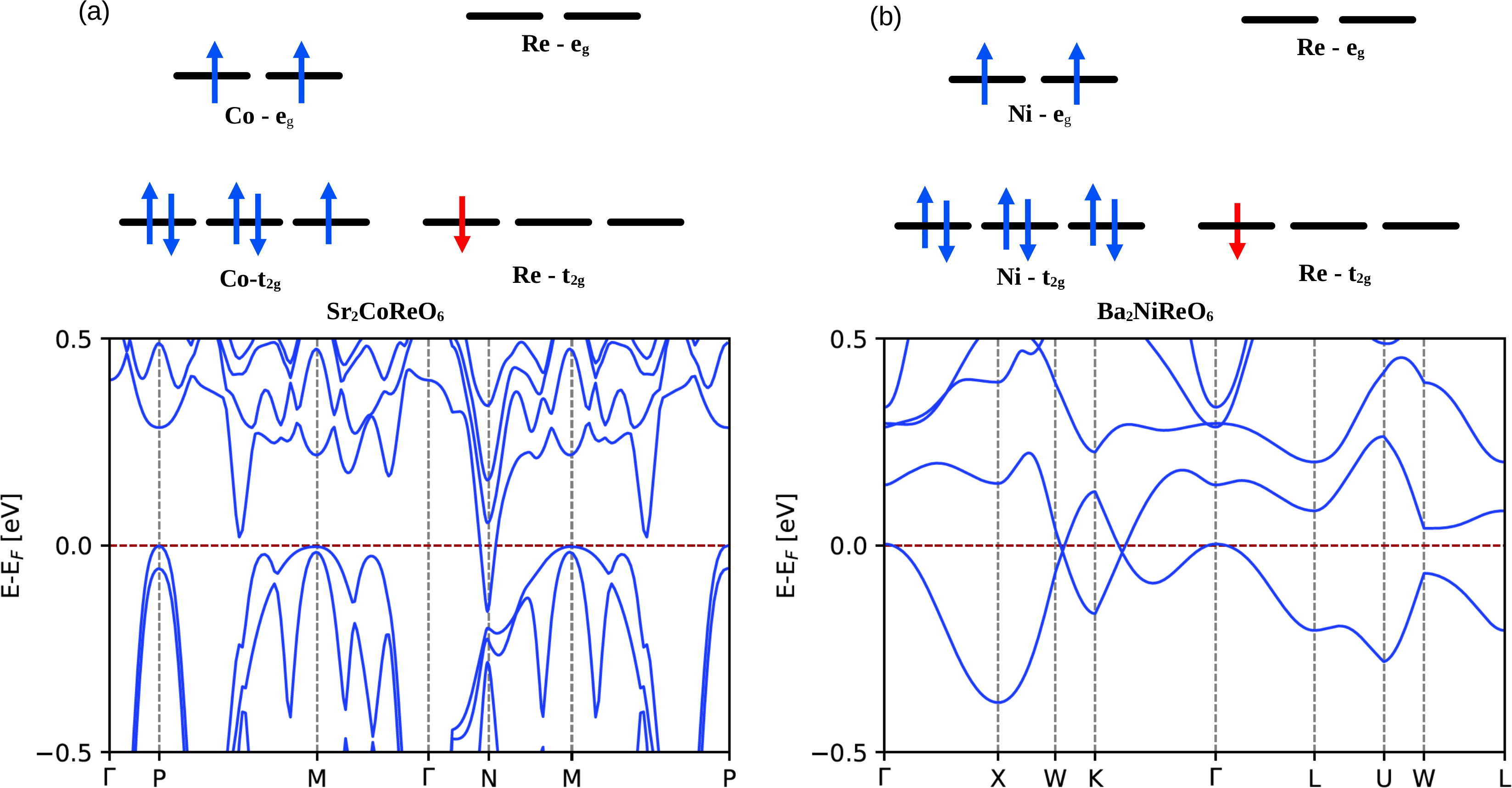}
		\caption{Schematic energy level diagram, orbital filling of the d levels in the octahedron crystal field at the Co/Ni-3d and Re-5d sites and band structure with SOC for (a) tetragonal double perovskites, Sr$_2$CoReO$_6$, and (b) cubic double perovskites, Ba$_2$NiReO$_6$. }
		\label{exchange-splitting}
	\end{figure*}

 Our calculations indicated that whenever B-3d magnetic sites have fully half-filled d-shells or half-filled e$_g$ states, because of the enhanced exchange splitting, the filled 3d-orbitals are pushed below the Fermi energy, significantly reducing mixing with the occupied  4d/5d-t$_{2g}$ states. As a result, a clean band structure with predominant 4d/5d-t$_{2g}$ orbital character is observed around the Fermi energy. For example, in Fig. \ref{exchange-splitting}(a,b), we present the band structure of  Sr$_2$CoReO$_6$ and Ba$_2$NiReO$_6$ with electron occupancy at the B and B$'$-sites of Co-d$^7$, Ni-d$^8$ and Re-d$^{1}$, respectively. Owing to the completely half-filled Ni-e$_g$ states in the compound, we observed a clean band structure with predominantly partially filled Re-t$_{2g}$ states near the Fermi energy compared to the Sr$_2$CoReO$_6$ compound. 
	
In the following section, we briefly describe the charge state and electron  occupancies of the d-orbitals for the considered compounds.  In the Fe based cubic DP, we find a magnetic moment of around ~4.0 $\mu_B$, indicating a nominal valency of Fe$^{3+}$ with a high-spin state d$^5$ occupancy [t$_{2g}$(3$\uparrow$), e$_g$(2$\uparrow$)] consistent with the experimental result reported for the Fe based compounds \cite{Fe-based-DP-exp1,Fe-based-DP-exp2,Fe-based-DP-exp3}. On the other hand, the calculated density of states (DOS) and magnetic moment indicated a nominal valency of "2+"  at the Mn, Co, and Ni with the following occupancies for d$^{5}$ [t$_{2g}$(3$\uparrow$), e$_g$(2$\uparrow$)], d$^{7}$ [t$_{2g}$(3$\uparrow$,2$\downarrow$), e$_g$(2$\uparrow$)] and d$^{8}$ [t$_{2g}$(3$\uparrow$, 3$\downarrow$), e$_g$(2$\uparrow$)], respectively. The d-level occupancy at the 4d/5d-B$'$ site of Mo, W, and Re varies from 0 to 2 (cf. Table I).  The low energy state is primarily governed by the 4d/5d-t$_{2g}$ states. So it is expected to have a negligible contribution from the 4d/5d-t$_{2g}$ orbital near the Fermi energy for the completely unoccupied 4d/5d-t$_{2g}$ state. Therefore, compounds with half-filled 3d-B orbitals  and completely unoccupied 4d/5d-t$_{2g}$ orbitals usually lead to an insulating solution as found in Ba$_2$MnMoO$_6$ and Ba$_2$NiWO$_6$.

In the tetragonal group of DP compounds, from the calculated DOS and magnetic moment, the sites occupied by Co, Ni and Fe, Cr are found to be in the  "2+" and "3+" charge state, respectively, with occupancies of  d$^{7}$ [t$_{2g}$(3$\uparrow$,2$\downarrow$), e$_g$(2$\uparrow$)], d$^{8}$ [t$_{2g}$(3$\uparrow$, 3$\downarrow$), e$_g$(2$\uparrow$)], d$^{5}$ [t$_{2g}$(3$\uparrow$), e$_g$(2$\uparrow$)], and d$^{3}$ [t$_{2g}$(3$\uparrow$)], respectively. In the group of tetragonal DP compounds, Sr$_2$NiMoO$_6$ and Sr$_2$CrOsO$_6$ were found to be insulating. In the Sr$_2$CrOsO$_6$ compound, both Cr and Os are in the d$^3$ configuration and aligned anti-parallel to each other. Both completely filled, the Cr-t$_{2g}$ up spin channel and Os-t$_{2g}$ down spin channel lie below the Fermi energy and the empty e$_g$ states of Cr and Os lie above the Fermi energy.  This leads to an insulating solution where as in the Sr$_2$NiMoO$_6$ compound Mo is the d$^0$ state, thus a negligible contribution around the Fermi energy. The half filled Ni-e$_g$ states lie below the Fermi energy and give rise to an insulating solution.

Based on the knowledge we gained of the electronic structure for all the considered DP compounds, we subsequently investigated the topological transport characteristics, AHC and ANC.  Table I summarizes the calculated AHC and ANC for all the compounds at the Fermi energy and supplementary Figure S1 shows a graphic representation. It is clearly evident that the value of AHC and ANC around the Fermi energy is more or less of the same order of magnitude for the cubic and tetragonal group of DP compounds, for example the cubic DP
Ba$_2$NiReO$_6$  and tetragonal DP Sr$_2$NiReO$_6$. 
To correlate the similar trend that was observed for the AHC and ANC in the cubic and tetragonal group of DP compounds, we investigated their magnetic space group in detail using the  FINDSYM \cite{findsym1,findsym2} software package. Considering
the ferrimagnetic alignment of the B-3d and B$'$-4d/5d sites, the magnetic space groups (MSG) of the cubic and tetragonal DP compounds were determined to be I4/m m$'$m$'$ and I4/m , respectively. Both MSG have a mirror m$_z$ along the c axis, and lack m$_x$ and m$_y$, which is why both cubic and tetragonal DPs present a gapless nodal line at k$_z$ = 0 but gapped  nodal lines at k$_x$ = 0 and k$_y$ = 0. However, these gapped nodal lines have a different origin, 
in the case of cubic DPs m$_x$ and m$_y$ are broken because of the coupling between SOC and magnetism, whereas in the case of tetragonal DPs they were never present. Hence, for the cubic DP compounds, the main source of the Berry curvature around the Fermi energy is found to be the gapped topological nodal line in the presence of SOC and is discussed in more detail in section A.

The calculated value of the AHC of the  Ba$_2$NiReO$_6$
is almost three times larger than that of the ferromagnetic SrRuO$_3$,which is a prototype of ferromagnetic oxide materials and has been studied extensively \cite{SrRuO3-Science,SrRuO3-2d}. In addition, the determination of the energy dependence of the AHC and ANC is also important to account for the energy shifts away from the Fermi energy. Therefore, we additionally show the largest possible value in a range of 200 meV around the Fermi energy in Table 1.  
This occurs because minor oxygen nonstoichiometry is commonly found in perovskite oxides\cite{chen-ang-oxygen-nonstoichiometry, lima-magnetic, wei-tong-double-exchange} and has the effect of a small degree of doping and, consequently, a shift in the Fermi level.
The largest values of AHC and ANC we found are 571 S/cm and -2.83 A m$^{-1}$ K$^{-1}$ in Ba$_2$NiReO$_6$. These results show that changes in the Fermi energy can have a large influence on the anomalous transport coefficients, AHC, and ANC. To understand the large anomalous transport coefficients of  Ba$_2$NiReO$_6$, we investigated its electronic and magnetic structure and topological features in more detail, and present the results in the following section.


\begin{figure}[ht!]
		\centering
			\includegraphics[width=0.45\textwidth]{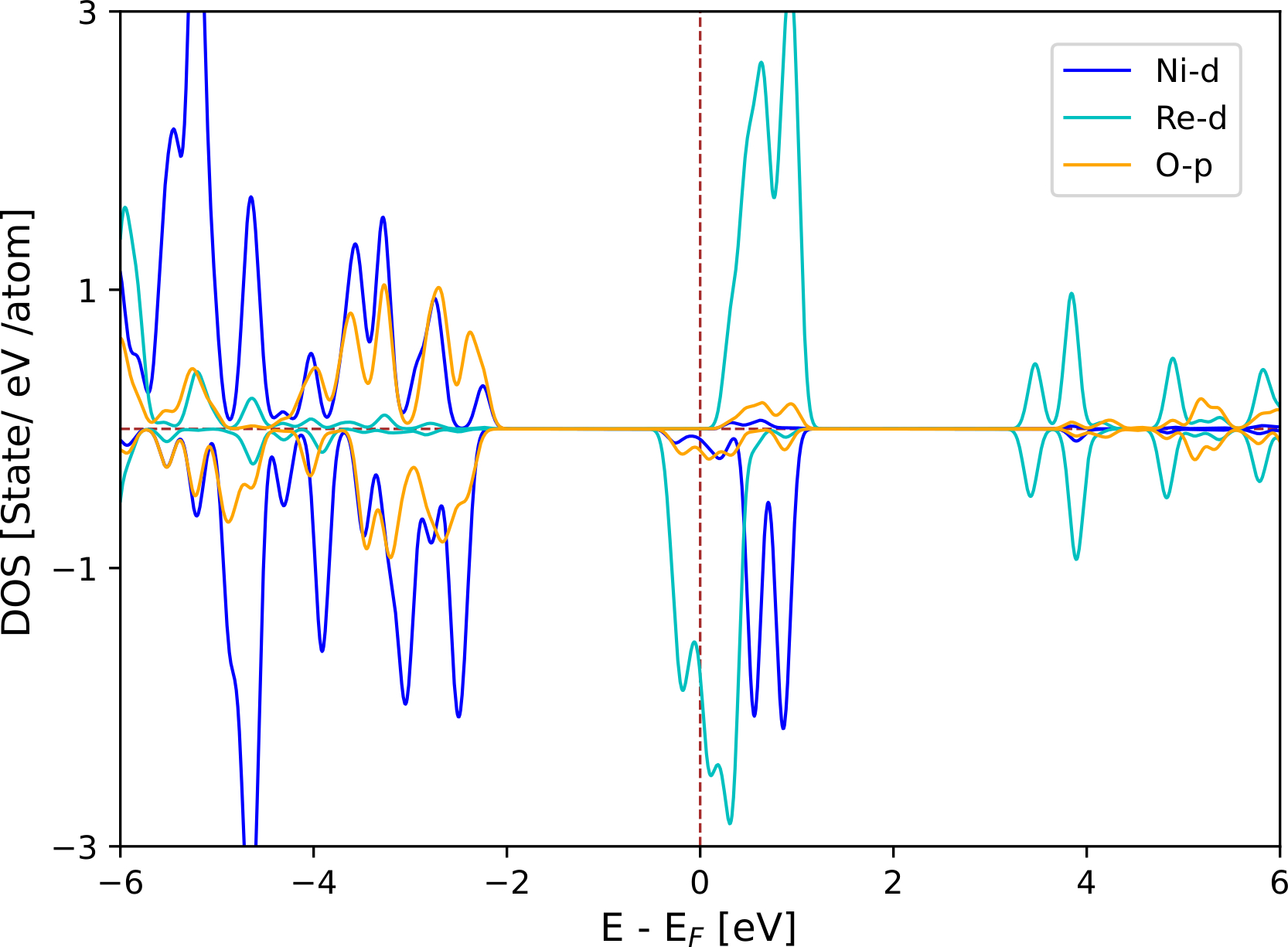}
		\caption{Spin-polarized GGA+$U$ density of states for the DP Ba$_2$NiReO$_6$, projected onto Ni $-$d (blue), Re$-$d, (cyan), and O$-$p (orange) states. Zero energy is specified to be equal to the GGA+$U4$ Fermi energy.}
		\label{dos}
	\end{figure}

\begin{figure*}[ht!]
		\centering
		\includegraphics[width=0.9\textwidth]{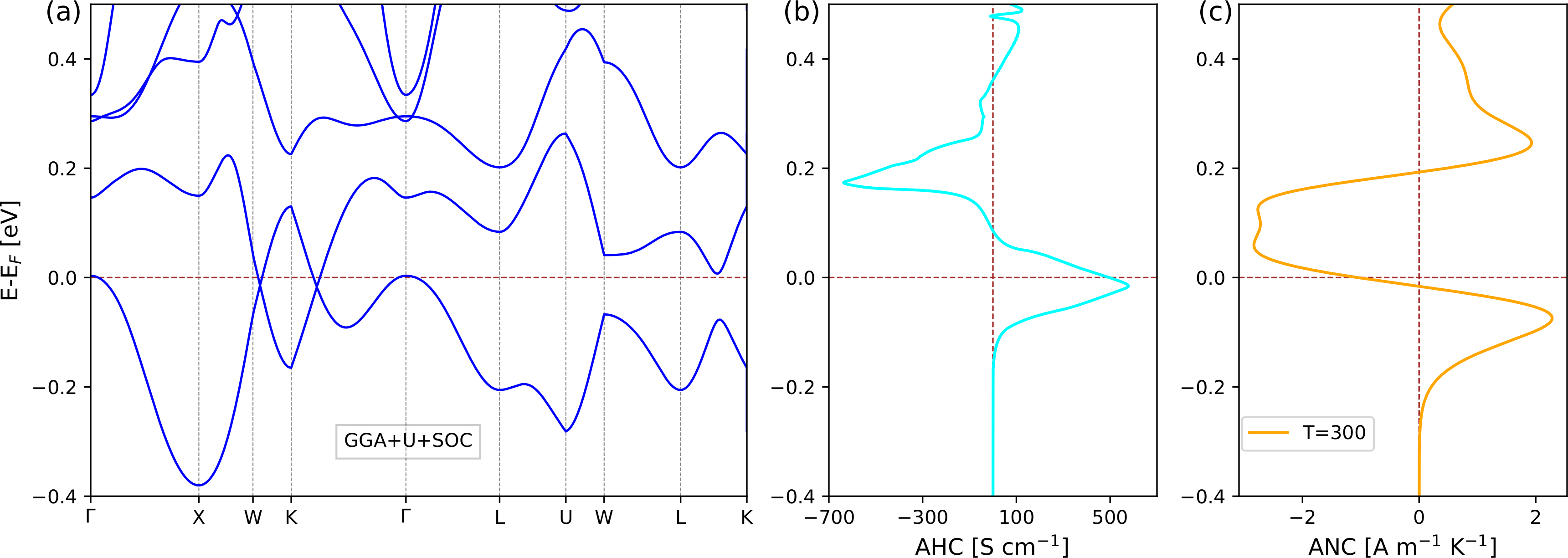}
		\caption{(a) Band structure in the presence of spin–orbit coupling (SOC) and Hubbard U. (b) anomalous Hall conductivity (AHC), $\sigma$,
 and (c) anomalous Nernst conductivity (ANC), $\alpha$, for the DP Ba$_2$NiReO$_6$.}
		\label{AHC-ANE}
	\end{figure*}

	\begin{figure*}[ht!]
		\centering
		\includegraphics[width=0.9\textwidth]{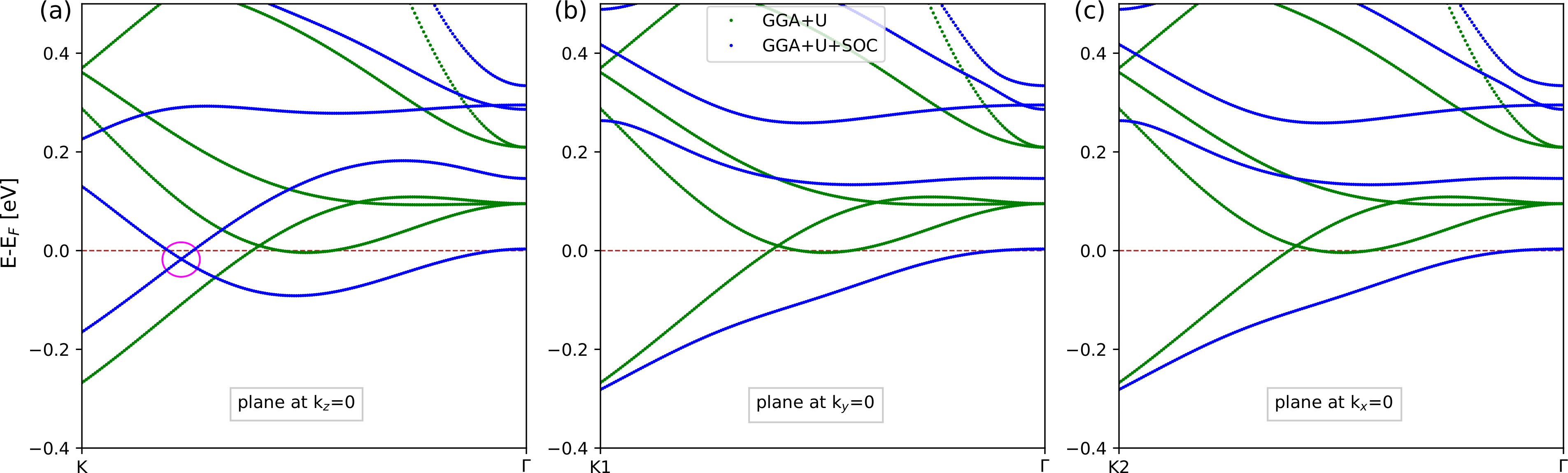}
		\caption{Band structure calculated along the $ K - \Gamma $ direction in the planes k$_{z}$ = 0 (a), k$_{y}$ = 0 (b), and k$_{x}$ = 0 (c) for the DP Ba$_2$NiReO$_6$. Green lines: band structure with GGA+$U$, blue lines: band structure with GGA+U+SOC. 
		In the k$_{z}$ = 0 plane, the crossing indicated by the red circle is intact due to the protected mirror plane while in the k$_x$ = 0 or k$_y$ = 0 planes, nodal lines gapped out in the presence of SOC (magnetization, M $\parallel$ (001).}
		\label{BMRO-nodal-line}
	\end{figure*}

	\begin{figure*}[ht!]
		\centering
		\includegraphics[width=0.9\textwidth]{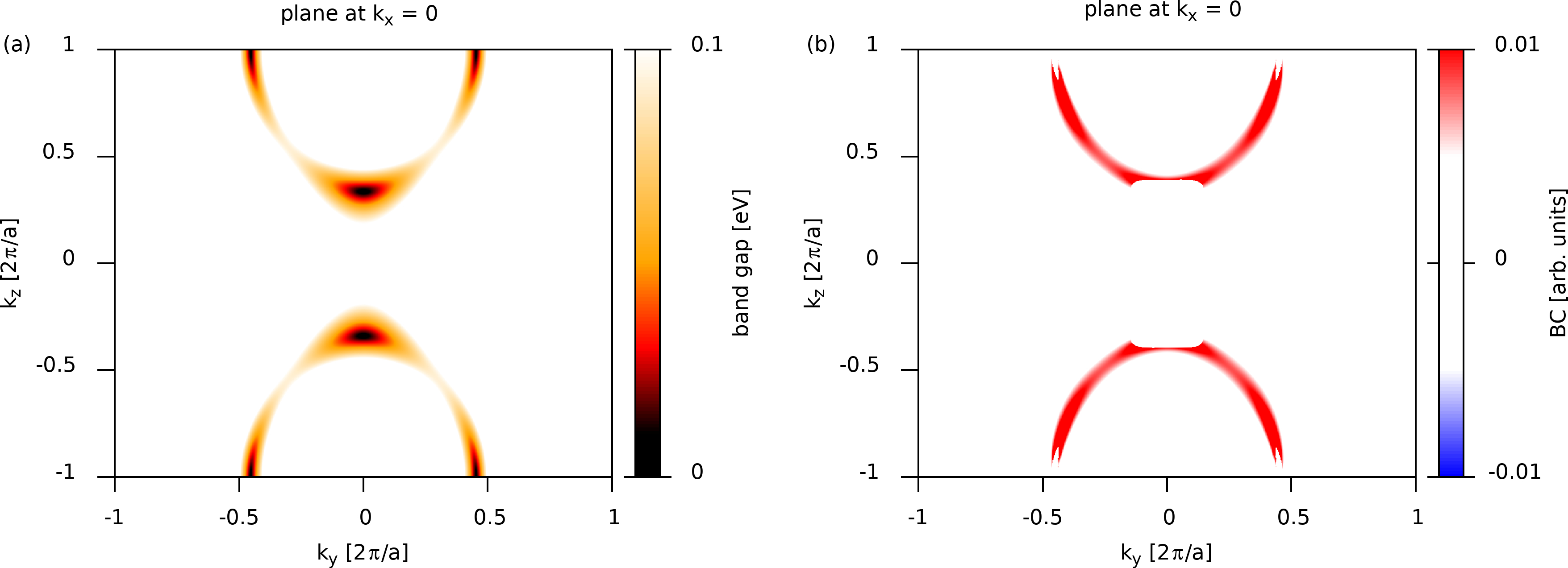}
		\caption{(a) Band gap in the k$_{x}$ = 0 plane with magnetization, M$\parallel$ (001). The Gaped nodal line is visible. (b) Berry curvature in the k$_{x}$ = 0 plane with M$\parallel$ (001). Strong Berry curvature contribution is visible along the gapped nodal line.}
		\label{BNRO-nodal-line}
	\end{figure*}

\subsection{Electronic structure and magnetic ground state of Ba$_2$NiReO$_6$ }

First, using the GGA+$U$ method, we compared the total energies of the different magnetic structures $-$ non-magnetic, ferromagnetic, and ferrimagnetic $-$ to determine the magnetic ground state of DP compounds. The calculations revealed the ferrimagnetic ground state to be energetically favorable. The calculated spin polarized DOS, considering the ferrimagnetic ground state with the GGA+$U$ calculation, is shown in Fig. \ref{dos}. 
Within the crystal field of a regular octahedron, the d-levels split into t$_{2g}$ and e$_g$ states and the d-states of Ni and the Re ions are exchanged and the crystal field splits. The exchange splitting at the 3d-Ni site is expected to be high compared to the 5d-Re site, whereas the large crystal field splitting at the 5d-Re site stabilizes in the low spin state. 
The calculations with the GGA+$U$ showed that  the localized magnetic moment at the Ni$-3d$ and Re$-5d$ sites is found to be 1.70 and -0.7 $\mu_{B}$, respectively, which indicates that Ni$^{2+}$(3d$^8$) is in the high spin ([t$_{2g}$(3$\uparrow$, 3$\downarrow$) e$_g$(2$\uparrow$)]) and Re$^{6+}$(5d$^1$) in the low spin  [t$_{2g}$(1$\downarrow$)] states.
 In the DP Ba$_2$NiReO$_6$ compound, as Ni is in the high spin d$^{8}$ configuration, both the majority and minority spin channels of t$_{2g}$  and the majority spin channel of e$_{g}$ are completely filled, whereas the minority spin channel of e$_{g}$ remains completely empty and lies above the Fermi energy (cf. Fig. \ref{dos}). On the other hand, low spin Re$^{6+}$(5d$^1$) has only one electron and, from Fig. \ref{dos}, it is evident that, the states close to the Fermi level (E$_F$) are mainly contributed by the partially filled Re-[t$_{2g}$(1$\downarrow$)] states in the down spin channel. 
 The partially filled Re-t$_{2g}$ states cross the Fermi level in the minority spin channel, which causes this spin channel to become conducting. In contrast, a clear gap arises in the majority spin channel, leading to a half-metallic ferrimagnetic ground state with full spin polarization. 
From the orbital-resolved partial DOS (Fig. \ref{dos}), it is clearly evident that a trivial band gap of 1.2 eV exists in the majority spin channel of  Ba$_2$NiReO$_6$, whereas that in the minority spin channel contributes significant states to form peak-like DOS near the Fermi level, which consist mainly of Re-t$_{2g}$ states hybridized with O-2p states.

As the states close to the Fermi energy are dominated by the partially filled Re-t$_{2g}$ states, the effect of the SOC in the electronic structure is expected to be high. The electronic band structure, with the GGA+$U$+SOC calculations considering the ferrimagnetic ground state, is shown in the Fig. \ref{AHC-ANE}a. With the constructed tight binding Hamiltonian considering the atomic orbital-like MLWFs of the Ni$-d$ , Re$-d$, Sr$-d$  and O$-p$ states, the calculated intrinsic contribution of the AHC and the ANC as a function of the Fermi energy is shown in the Fig. \ref{AHC-ANE}(b,c), respectively. 

From the computed value of the AHC as a function of Fermi energy, we observed a large peaks of the AHC around the Fermi energy (cf. Fig. \ref{AHC-ANE}b).
We also crosschecked our electronic structure, AHC and ANC calculations for a different choice of $U$ values at the Ni$-3d$ and Re$-5d$ site, but no significant changes were observed in the basic electronic structure and in the peak of AHC and ANC near the Fermi energy (cf. Supplementary Figure S2). This demonstrates the robustness of our calculation. In the next section, we investigated the peak in the AHC in detail, taking into account the values of U $=$ 4.0 and 1.6 eV at Ni$-3d$ and Re$-5d$ sites, respectively.
 
\begin{figure*}[ht!]
		\centering
			\includegraphics[width=0.9\textwidth]{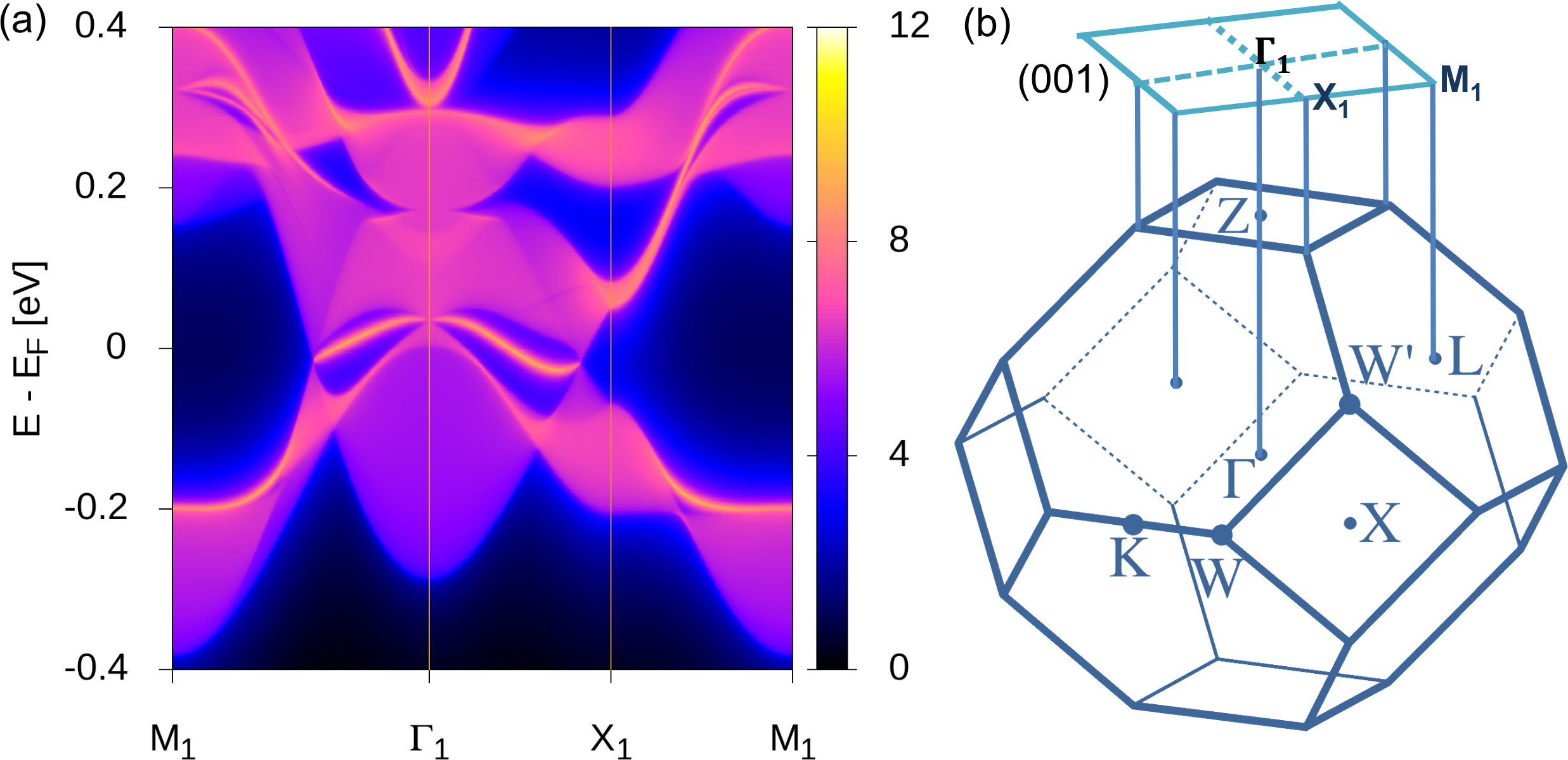}
		\caption{(a) Band dispersion of Ba$_2$NiReO$_6$ on the projected (001) surface. (b) Brillouin zones (BZs) of bulk and (001) surfaces with high-symmetry points of the fcc structure.}
		\label{surface}
	\end{figure*}

\subsection{Topological analysis of Ba$_2$NiReO$_6$}

 The band structure of the DP Ba$_2$NiReO$_6$ undergoes a very distinct change because of the strong SOC of the Re$-$5d states. Near the Fermi level, most of the band touching points are gapped. Along the $K-\Gamma$ directions, the valence and conduction bands of Re-$t_{2g}$ intersect with each other. To understand the topological nature of the band crossing along $K-\Gamma$, we examined the electronic structure in more detail.
In the absence of magnetism and the SOC, the cubic DP Ba$_2$NiReO$_6$ possesses m$_x$, m$_y$, m$_z$, C$_{4x}$, C$_{4y}$, and C$_{4z}$ symmetries. 
Without SOC, the internal spin-space, and the lattice space are decoupled from each other and spins can be separated into spin-up and spin-down sectors. Thus, in the absence of the SOC, cubic Ba$_2$NiReO$_6$ possesses three mirror planes at k$_{z}$ = 0, k$_{y}$ = 0, k$_{x}$ = 0 and symmetry-protected nodal lines can appear in any of the three planes \cite{nodal1,nodal2,nodal3}. Considering the magnetic ground state with GGA+$U$, we observe the symmetry-protected nodal line in each of the three planes (k$_x$ = 0, k$_y$ = 0, k$_z$ = 0) along the $K-\Gamma$ direction, as shown with green lines in Fig. \ref{BMRO-nodal-line}(a,b,c).

The situation is different when we take into account the SOC, in the presence of which the spin-space (coupled up- and down-spin channels) and the lattice space are coupled with each other, thus breaking the spin-rotational symmetry. Particularly, when considering the moments aligned along the z$-$direction, i.e., the SOC along the 001-direction, the m$_x$ and m$_y$ symmetries are broken, thus breaking the symmetry protection of nodal lines in the k$_x$ = 0 and k$_y$ = 0 planes, leading to a gapped nodal line along the $K-\Gamma$ high symmetry line  (cf. Fig.  \ref{BMRO-nodal-line}(c,b)). Instead, for the moment aligned along the z-direction, the m$_z$ and C$_{4z}$ symmetries are still preserved, which leads to the gapless nodal line in the k$_z$ = 0 plane (cf. Fig. \ref{BMRO-nodal-line}a). As a result of the gapped nodal lines, a large Berry curvature could be induced in the Brillouin zone (BZ), consequently enhancing the AHC\cite{nodal1,nodal2,nodal3}.

We calculated the Berry curvature contribution in these planes to further investigate the contribution of these gapped nodal lines to the AHC.
As expected, due to the presence of the mirror symmetry at  the $k_{z}$ = 0 plane with SOC (M $\parallel$ 001), a closed nodal line would be visible and, consequently,  the closed nodal line results in an almost zero Berry curvature contribution in the $k_z$ = 0 plane.
In the Fig. \ref{BNRO-nodal-line}a, we show the band gap in the $k_x$ = 0 plane by considering the two bands that form the nodal line for the DP Ba$_2$NiReO$_6$. In the $k_x$ = 0 plane, in which the mirror symmetry is broken due to the magnetization along the z-direction (001), the nodal lines are separated by a gap, as shown in Fig. \ref{BNRO-nodal-line}a. Consequently, the gapped nodal lines make a large contribution to the Berry curvature in the system, as shown in Fig. \ref{BNRO-nodal-line}b. To quantify the contributions that the gapped nodal lines make to the total AHC, we restricted the integration of Eq. \ref{AHC} to slices around the $k_x$ = 0 and $k_y$ = 0 planes, as these are the planes in which the gapped lines reside.  Considering the thickness of the slices at 20$\%$ of the full BZ, 
we observe that 80 $\%$ of the AHC value can be attributed to the gapped nodal line structure of the DP Ba$_2$NiReO$_6$, which suggests that the gapped nodal lines are the primary source of AHC.


\subsection{Topological surface state}

One of the hallmark features of the nodal line compounds are the protected drumhead surface states, consisting of surface states covering the projection of the nodal line on the surface. Therefore, to find the topologically protected surface state of DP Ba$_2$NiReO$_6$, the surface states were calculated using the WannierTools software package \cite{WanTool1, WanTool2}, which employs tight-binding Hamiltonians constructed by maximally localized Wannier functions \cite{wan1,wan4}.
In the topological nodal line compounds, the topological invariants \cite{Nodal-invariant} and, thus, the surface spectrum are protected by crystalline symmetries. Therefore  only surfaces which preserve them will show topologically protected surface states\cite{Nodal-sur1,Nodal-sur2,Nodal-sur3}. In the case of the DP Ba$_2$NiReO$_6$, the mirror symmetry protected nodal line is observed in the plane k$_z$ = 0, thus one could expect a surface state to prevail on the  (001) surface. 

In Fig. \ref{surface}a  we show the projection of the band dispersion onto the (001) surface, following the high-symmetry k paths shown in the Fig. \ref{surface}b. On the projected (001) surface, topological nontrivial surface states are clearly visible, connecting the two gapless points within $\Gamma_{1}-X_{1}$ and $M_1- \Gamma_{1}$ to form a two-dimensional drumhead surface.
Moreover, the dispersion of the drumhead surface states are observed to be small and located very close to the Fermi energy, contributing a large surface DOS at the Fermi energy, which is greatly beneficial for experimental realization as well as in spintronics applications\cite{Nodal-sur-application}.

	
\section{Summary}

Using the first-principles density functional, we investigated the topological transport characteristics, AHC and ANC for stable cubic and tetragonal rock-salt ordered DPs (A$_2$BB$'$O$_6$) considering the 3d TM magnetic ions at the B site and 4d/5d TM magnetic ions at the B$'$ site. 
Using first-principles DFT calculations and Wannier interpolation, we have investigated the topological properties, AHC and ANC of a list of representative materials. Our calculations of the electronic structure showed that DP compounds with a half-filled B-site 3d-t$_{2g}$ or e$_{g}$ and partially filled 4d/5d-t$_{2g}$ states, lead to the possibility of a clean topological band crossing. This is primarily attributed to strong SOC in the 5d-t$_{2g}$ orbital near the Fermi energy with large AHC.

 
 The magnetic space group of the cubic and tetragonal DP compounds considering the ferri-magnetic aligned of B-3d and B$'$-4d/5d site, is found to be in the I4/m m$'$m$'$ and I4/m , respectively, with a mirror symmetry m$_z$ along the c axis in the $k_z$ = 0 plane. Prior to the inclusion of SOC and magnetism, the cubic structure also had $m_x$ and $m_y$ mirror symmetries, this is why the primary source of the Berry curvature for the cubic DP compounds around the Fermi energy was to be the gapped topological nodal line in the $k_x$=0 and $k_y$=0 planes. The presence of mirror symmetries and a clean topological band crossing primarily resulting from the 5d-t$_{2g}$ state, were found to be crucial for large AHC and ANC, which, in combination with the magnetism, produces a large BC in the band structure. In certain DP compounds, particularly in Ba$_2$NiReO$_6$, we
found a symmetry-protected nodal line 10 meV below the Fermi energy with very large AHC and/or ANC values that are three times larger than the extensively studied prototype metallic ferromagnetic oxide material SrRuO$_3$ \cite{SrRuO3-Science,SrRuO3-2d}. 
 In the DP Ba$_2$NiReO$_6$, slightly dispersing mirror symmetry protected drumhead surface states is observed around the Fermi energy making it an excellent candidates for using the topological properties in the spintronics applications. We additionally found that both AHC and ANC are strongly dependent on the position of the Fermi level and therefore on the doping level of the investigated material. This work demonstrates the versatility and usability of double perovskites based on 3d and 4d/5d TM ions to realize a clean topological band crossing primarily resulting from the  4d/5d-t$_{2g}$ state with a large AHC. Our findings pave the way for new high-performance stable magnetic topological materials based on oxide double perovskites.

	\section{Acknowledgements}
This work is financially supported by the European Research Council (ERC Advanced Grant No. 742068 ‘TOPMAT’). We also acknowledge funding by the DFG through SFB 1143 (project ID. 247310070) and the W\"{u}rzburg-Dresden Cluster of Excellence on Complexity and Topology in Quantum Matter ct.qmat (EXC2147, project ID. 39085490). M.G.V. and I.R. thanks support from the Spanish Ministerio de Ciencia e Innovacion (grant PID2019-109905GB-C21) and European Research Council (ERC) grant agreement no. 101020833.
M.G.V. and C.F. acknowledge support by the Deutsche Forschungsgemeinschaft (DFG, German Research Foundation) – FOR 5249 (QUAST).



\end{document}